\documentclass[aps,prd,showpacs,twocolumn,superscriptaddress]{revtex4}  % for submission

\usepackage{graphicx}  % needed for figures
\usepackage{dcolumn}   % needed for some tables
\usepackage{bm}        % for math
\usepackage{amssymb}   % for math
\usepackage{epsfig}
\RequirePackage[usenames]{pstcol}

\def\met{{\mbox{$E\kern-0.57em\raise0.19ex\hbox{/}_{T}$}}}

\newcommand{\ttbar}{\mbox{$t{\bar t}~$}}

\begin{document}
\leftline{ }
\rightline{FERMILAB-PUB-10-125-E}

%\rightline{CDF Note 10101, D0 Note 6039}

\title{Combined Tevatron upper limit on {\boldmath $gg\rightarrow H\rightarrow W^+W^-$}\\
and constraints on the Higgs boson mass in fourth-generation fermion models}
%and constraints on fourth-generation fermion models}

\date{\today}

\begin{abstract}

\vskip 1cm
We combine results from searches by the CDF and D0 collaborations for a standard model Higgs
boson ($H$) in the process $gg\rightarrow H\rightarrow W^+W^-$ in $p{\bar{p}}$ 
collisions at the Fermilab Tevatron Collider at
$\sqrt{s}=1.96$~TeV.  With 4.8~fb$^{-1}$ of integrated luminosity analyzed at CDF and
5.4~fb$^{-1}$ at D0, the 95\% Confidence Level upper limit on $\sigma(gg\rightarrow H)\times \mathcal{B}(H\rightarrow W^+W^-)$ 
is 1.75 pb at $m_H=120$~GeV, 0.38 pb at $m_H=165$~GeV, and 0.83 pb at $m_H=200$~GeV.  Assuming
the presence of a fourth sequential generation of fermions with large masses,
 we exclude at the 95\% Confidence Level a standard-model-like Higgs boson with a mass between 131 and 204~GeV.
\end{abstract}
%14.80.Bn 	Standard-model Higgs bosons
%14.70.Fm 	W bosons
%13.85.Rm 	Limits on production of particles 
%14.65.Jk 	Other quarks (e.g., 4th generations) 

\affiliation{Universidad de Buenos Aires, Buenos Aires, Argentina}
\affiliation{LAFEX, Centro Brasileiro de Pesquisas F{\'\i}sicas, Rio de Janeiro, Brazil}
\affiliation{Universidade do Estado do Rio de Janeiro, Rio de Janeiro, Brazil}
\affiliation{Universidade Federal do ABC, Santo Andr\'e, Brazil}
\affiliation{Instituto de F\'{\i}sica Te\'orica, Universidade Estadual Paulista, S\~ao Paulo, Brazil}
\affiliation{Institute of Particle Physics: McGill University, Montr\'{e}al, Qu\'{e}bec, Canada; Simon Fraser University, Burnaby, British Columbia, Canada; University of Toronto, Toronto, Ontario, Canada; and TRIUMF, Vancouver, British Columbia, Canada}
\affiliation{Simon Fraser University, Burnaby, British Columbia, Canada; and York University, Toronto, Ontario, Canada}
\affiliation{University of Science and Technology of China, Hefei, People's Republic of China}
\affiliation{Institute of Physics, Academia Sinica, Taipei, Taiwan, Republic of China}
\affiliation{Universidad de los Andes, Bogot\'{a}, Colombia}
\affiliation{Charles University, Faculty of Mathematics and Physics, Center for Particle Physics, Prague, Czech Republic}
\affiliation{Czech Technical University in Prague, Prague, Czech Republic}
\affiliation{Center for Particle Physics, Institute of Physics, Academy of Sciences of the Czech Republic, Prague, Czech Republic}
\affiliation{Universidad San Francisco de Quito, Quito, Ecuador}
\affiliation{Division of High Energy Physics, Department of Physics, University of Helsinki and Helsinki Institute of Physics, FIN-00014, Helsinki, Finland}
\affiliation{LPC, Universit\'e Blaise Pascal, CNRS/IN2P3, Clermont, France}
\affiliation{LPSC, Universit\'e Joseph Fourier Grenoble 1, CNRS/IN2P3, Institut National Polytechnique de Grenoble, Grenoble, France}
\affiliation{CPPM, Aix-Marseille Universit\'e, CNRS/IN2P3, Marseille, France}
\affiliation{LAL, Universit\'e Paris-Sud, CNRS/IN2P3, Orsay, France}
\affiliation{LPNHE, Universit\'es Paris VI and VII, CNRS/IN2P3, Paris, France}
\affiliation{CEA, Irfu, SPP, Saclay, France}
\affiliation{IPHC, Universit\'e de Strasbourg, CNRS/IN2P3, Strasbourg, France}
\affiliation{IPNL, Universit\'e Lyon 1, CNRS/IN2P3, Villeurbanne, France and Universit\'e de Lyon, Lyon, France}
\affiliation{III. Physikalisches Institut A, RWTH Aachen University, Aachen, Germany}
\affiliation{Physikalisches Institut, Universit{\"a}t Bonn, Bonn, Germany}
\affiliation{Physikalisches Institut, Universit{\"a}t Freiburg, Freiburg, Germany}
\affiliation{II. Physikalisches Institut, Georg-August-Universit{\"a}t G\"ottingen, G\"ottingen, Germany}
\affiliation{Institut f\"{u}r Experimentelle Kernphysik, Karlsruhe Institute of Technology, Karlsruhe, Germany}
\affiliation{Institut f{\"u}r Physik, Universit{\"a}t Mainz, Mainz, Germany}
\affiliation{Ludwig-Maximilians-Universit{\"a}t M{\"u}nchen, M{\"u}nchen, Germany}
\affiliation{Fachbereich Physik, Bergische Univerit{\"a}t Wuppertal, Wuppertal, Germany}
\affiliation{University of Athens, 157 71 Athens, Greece}
\affiliation{Panjab University, Chandigarh, India}
\affiliation{Delhi University, Delhi, India}
\affiliation{Tata Institute of Fundamental Research, Mumbai, India}
\affiliation{University College Dublin, Dublin, Ireland}
\affiliation{Istituto Nazionale di Fisica Nucleare Bologna, $^{ee}$University of Bologna, I-40127 Bologna, Italy}
\affiliation{Laboratori Nazionali di Frascati, Istituto Nazionale di Fisica Nucleare, I-00044 Frascati, Italy}
\affiliation{Istituto Nazionale di Fisica Nucleare, Sezione di Padova-Trento, $^{ff}$University of Padova, I-35131 Padova, Italy}
\affiliation{Istituto Nazionale di Fisica Nucleare Pisa, $^{gg}$University of Pisa, $^{hh}$University of Siena and $^{ii}$Scuola Normale Superiore, I-56127 Pisa, Italy}
\affiliation{Istituto Nazionale di Fisica Nucleare, Sezione di Roma 1, $^{jj}$Sapienza Universit\`{a} di Roma, I-00185 Roma, Italy}
\affiliation{Istituto Nazionale di Fisica Nucleare Trieste/Udine, I-34100 Trieste, $^{kk}$University of Trieste/Udine, I-33100 Udine, Italy}
\affiliation{Okayama University, Okayama 700-8530, Japan}
\affiliation{Osaka City University, Osaka 588, Japan}
\affiliation{University of Tsukuba, Tsukuba, Ibaraki 305, Japan}
\affiliation{Waseda University, Tokyo 169, Japan}
\affiliation{Center for High Energy Physics: Kyungpook National University, Daegu, Korea; Seoul National University, Seoul, Korea; Sungkyunkwan University, Suwon, Korea; Korea Institute of Science and Technology Information, Daejeon, Korea; Chonnam National University, Gwangju, Korea; Chonbuk National University, Jeonju, Korea}
\affiliation{Korea Detector Laboratory, Korea University, Seoul, Korea}
\affiliation{SungKyunKwan University, Suwon, Korea}
\affiliation{CINVESTAV, Mexico City, Mexico}
\affiliation{FOM-Institute NIKHEF and University of Amsterdam/NIKHEF, Amsterdam, The Netherlands}
\affiliation{Radboud University Nijmegen/NIKHEF, Nijmegen, The Netherlands}
\affiliation{Joint Institute for Nuclear Research, Dubna, Russia}
\affiliation{Institute for Theoretical and Experimental Physics, Moscow, Russia}
\affiliation{Moscow State University, Moscow, Russia}
\affiliation{Institute for High Energy Physics, Protvino, Russia}
\affiliation{Petersburg Nuclear Physics Institute, St. Petersburg, Russia}
\affiliation{Comenius University, 842 48 Bratislava, Slovakia; Institute of Experimental Physics, 040 01 Kosice, Slovakia}
\affiliation{Institut de Fisica d'Altes Energies, Universitat Autonoma de Barcelona, E-08193, Bellaterra (Barcelona), Spain}
\affiliation{Centro de Investigaciones Energeticas Medioambientales y Tecnologicas, E-28040 Madrid, Spain}
\affiliation{Instituto de Fisica de Cantabria, CSIC-University of Cantabria, 39005 Santander, Spain}
\affiliation{Stockholm University, Stockholm, Sweden, and Uppsala University, Uppsala, Sweden}
\affiliation{University of Geneva, CH-1211 Geneva 4, Switzerland}
\affiliation{Glasgow University, Glasgow G12 8QQ, United Kingdom}
\affiliation{Lancaster University, Lancaster LA1 4YB, United Kingdom}
\affiliation{University of Liverpool, Liverpool L69 7ZE, United Kingdom}
\affiliation{Imperial College London, London SW7 2AZ, United Kingdom}
\affiliation{University College London, London WC1E 6BT, United Kingdom}
\affiliation{The University of Manchester, Manchester M13 9PL, United Kingdom}
\affiliation{University of Oxford, Oxford OX1 3RH, United Kingdom}
\affiliation{University of Arizona, Tucson, Arizona 85721, USA}
\affiliation{Ernest Orlando Lawrence Berkeley National Laboratory, Berkeley, California 94720, USA}
\affiliation{University of California, Davis, Davis, California 95616, USA}
\affiliation{University of California, San Diego, La Jolla, California 92093, USA}
\affiliation{University of California, Los Angeles, Los Angeles, California 90024, USA}
\affiliation{University of California Riverside, Riverside, California 92521, USA}
\affiliation{University of California, Santa Barbara, Santa Barbara, California 93106, USA}
\affiliation{Yale University, New Haven, Connecticut 06520, USA}
\affiliation{University of Florida, Gainesville, Florida 32611, USA}
\affiliation{Florida State University, Tallahassee, Florida 32306, USA}
\affiliation{Argonne National Laboratory, Argonne, Illinois 60439, USA}
\affiliation{Fermi National Accelerator Laboratory, Batavia, Illinois 60510, USA}
\affiliation{Enrico Fermi Institute, University of Chicago, Chicago, Illinois 60637, USA}
\affiliation{University of Illinois at Chicago, Chicago, Illinois 60607, USA}
\affiliation{Northern Illinois University, DeKalb, Illinois 60115, USA}
\affiliation{Northwestern University, Evanston, Illinois 60208, USA}
\affiliation{University of Illinois, Urbana, Illinois 61801, USA}
\affiliation{Indiana University, Bloomington, Indiana 47405, USA}
\affiliation{Purdue University Calumet, Hammond, Indiana 46323, USA}
\affiliation{University of Notre Dame, Notre Dame, Indiana 46556, USA}
\affiliation{Purdue University, West Lafayette, Indiana 47907, USA}
\affiliation{Iowa State University, Ames, Iowa 50011, USA}
\affiliation{University of Kansas, Lawrence, Kansas 66045, USA}
\affiliation{Kansas State University, Manhattan, Kansas 66506, USA}
\affiliation{Louisiana Tech University, Ruston, Louisiana 71272, USA}
\affiliation{The Johns Hopkins University, Baltimore, Maryland 21218, USA}
\affiliation{University of Maryland, College Park, Maryland 20742, USA}
\affiliation{Boston University, Boston, Massachusetts 02215, USA}
\affiliation{Northeastern University, Boston, Massachusetts 02115, USA}
\affiliation{Harvard University, Cambridge, Massachusetts 02138, USA}
\affiliation{Massachusetts Institute of Technology, Cambridge, Massachusetts 02139, USA}
\affiliation{Tufts University, Medford, Massachusetts 02155, USA}
\affiliation{Brandeis University, Waltham, Massachusetts 02254, USA}
\affiliation{University of Michigan, Ann Arbor, Michigan 48109, USA}
\affiliation{Wayne State University, Detroit, Michigan 48201, USA}
\affiliation{Michigan State University, East Lansing, Michigan 48824, USA}
\affiliation{University of Mississippi, University, Mississippi 38677, USA}
\affiliation{University of Nebraska, Lincoln, Nebraska 68588, USA}
\affiliation{Rutgers University, Piscataway, New Jersey 08855, USA}
\affiliation{Princeton University, Princeton, New Jersey 08544, USA}
\affiliation{University of New Mexico, Albuquerque, New Mexico 87131, USA}
\affiliation{State University of New York, Buffalo, New York 14260, USA}
\affiliation{Columbia University, New York, New York 10027, USA}
\affiliation{The Rockefeller University, New York, New York 10021, USA}
\affiliation{University of Rochester, Rochester, New York 14627, USA}
\affiliation{State University of New York, Stony Brook, New York 11794, USA}
\affiliation{Brookhaven National Laboratory, Upton, New York 11973, USA}
\affiliation{Duke University, Durham, North Carolina 27708, USA}
\affiliation{The Ohio State University, Columbus, Ohio 43210, USA}
\affiliation{Langston University, Langston, Oklahoma 73050, USA}
\affiliation{University of Oklahoma, Norman, Oklahoma 73019, USA}
\affiliation{Oklahoma State University, Stillwater, Oklahoma 74078, USA}
\affiliation{University of Pennsylvania, Philadelphia, Pennsylvania 19104, USA}
\affiliation{Carnegie Mellon University, Pittsburgh, Pennsylvania 15213, USA}
\affiliation{University of Pittsburgh, Pittsburgh, Pennsylvania 15260, USA}
\affiliation{Brown University, Providence, Rhode Island 02912, USA}
\affiliation{University of Texas, Arlington, Texas 76019, USA}
\affiliation{Texas A\&M University, College Station, Texas 77843, USA}
\affiliation{Southern Methodist University, Dallas, Texas 75275, USA}
\affiliation{Rice University, Houston, Texas 77005, USA}
\affiliation{Baylor University, Waco, Texas 76798, USA}
\affiliation{University of Virginia, Charlottesville, Virginia 22901, USA}
\affiliation{University of Washington, Seattle, Washington 98195, USA}
\affiliation{University of Wisconsin, Madison, Wisconsin 53706, USA}

\author{T.~Aaltonen$^{\dag}$} \affiliation{Division of High Energy Physics, Department of Physics, University of Helsinki and Helsinki Institute of Physics, FIN-00014, Helsinki, Finland}
\author{V.M.~Abazov$^{\ddag}$} \affiliation{Joint Institute for Nuclear Research, Dubna, Russia}
\author{B.~Abbott$^{\ddag}$} \affiliation{University of Oklahoma, Norman, Oklahoma 73019, USA}
\author{M.~Abolins$^{\ddag}$} \affiliation{Michigan State University, East Lansing, Michigan 48824, USA}
\author{B.S.~Acharya$^{\ddag}$} \affiliation{Tata Institute of Fundamental Research, Mumbai, India}
\author{M.~Adams$^{\ddag}$} \affiliation{University of Illinois at Chicago, Chicago, Illinois 60607, USA}
\author{T.~Adams$^{\ddag}$} \affiliation{Florida State University, Tallahassee, Florida 32306, USA}
\author{J.~Adelman$^{\dag}$} \affiliation{Enrico Fermi Institute, University of Chicago, Chicago, Illinois 60637, USA}
\author{E.~Aguilo$^{\ddag}$} \affiliation{Simon Fraser University, Burnaby, British Columbia, Canada; and York University, Toronto, Ontario, Canada}
\author{G.D.~Alexeev$^{\ddag}$} \affiliation{Joint Institute for Nuclear Research, Dubna, Russia}
% ...
%\footnote{(The CDF$^{\dag}$ and D0$^{\ddag}$ Collaborations)}

\collaboration{The CDF$^\dag$ and D0$^\ddag$ Collaborations}

\maketitle

%\newpage
%\linenumbers

Exploring the mechanism for breaking the $SU(2)\times U(1)$
electroweak gauge symmetry is a priority in high energy physics.
Not only are this symmetry and its breaking~\cite{higgs} necessary
components for the consistency of the successful standard model
(SM)~\cite{gws}, but measurable properties of the breaking mechanism
are also very sensitive to possible phenomena that have not yet been observed
at collider experiments.  Measuring these properties, or setting limits on them, 
can constrain broad classes of extensions to the SM.

A natural extension to the SM that can be tested with Higgs boson
search results at the Fermilab Tevatron Collider is the presence of a fourth generation of
fermions with masses much larger than those of the three known
generations~\cite{fourthgen}.  While fits to precision electroweak data favor
a low-mass Higgs boson in the SM, the addition of a fourth generation
of fermions to the SM modifies the fit parameters such that a heavy Higgs boson is consistent for
up to $m_H\approx 300$~GeV at the 68\% Confidence Level (C.L.)~\cite{g4_hdecay}.
Measurements of the $Z$
boson decay width~\cite{lepzpole} exclude models in which the
fourth neutrino mass eigenstate has a mass less than 45~GeV.
If the neutrino masses are very large, however, a fourth generation of fermions is
not yet excluded.

One consequence of the extra fermions is that the
$ggH$ coupling is enhanced by a factor of roughly three relative to the SM
coupling~\cite{arik,g4_hdecay,abf}.  Since the
lowest-order $ggH$ coupling arises from a quark loop.  The top
quark contribution is the largest due to its large coupling with the
Higgs boson.  In the limit $m_{q4}\gg m_H$, where $m_{q4}$ is the fourth-generation quark mass, the Higgs boson coupling
cancels the mass dependence for each of the three propagators in the
loop, and the contribution to the $ggH$ coupling becomes asymptotically
independent of the masses of the two fourth-generation quarks.  Each
additional fourth-generation quark then contributes as much as the top
quark, and the $ggH$ coupling is thus enhanced by a factor $K_e$ of
approximately three.

The production cross section will be enhanced by a factor of $K_e^2$.
For $m_H$ near the low end of our search range, $m_H \approx 110$~GeV, the $gg\rightarrow H$ production
cross section is enhanced by roughly a factor of nine relative to the SM prediction.  This factor
drops to approximately 7.5 near the upper end of the search range, $m_H \approx 300$~GeV, assuming
asymptotically large masses for the fourth-generation quarks.  The
reason for this drop is that the denominator of the enhancement factor, the SM cross section, has a larger contribution
 from the SM top quark as $m_H$ nears $2m_t$.
The partial decay width for $H\rightarrow gg$ is enhanced by the same factor as the production
cross section.  However, because the decay $H\rightarrow gg$ is loop-mediated, the $H\rightarrow W^+W^-$ decay
continues to dominate for Higgs boson masses $m_H>135$~GeV.

We consider two scenarios for the masses of the fourth-generation fermions.
In the first scenario, the ``low-mass'' scenario, we set the mass of the fourth-generation neutrino 
to $m_{\nu 4}=80$~GeV, and the mass of the fourth-generation charged lepton to $m_{\ell 4}=100$~GeV
in order to evade experimental constraints~\cite{L3_lepton} and to have the maximum impact on the Higgs boson
decay branching ratios.  In the second scenario, the ``high-mass'' scenario,
 we set $m_{\nu 4}=m_{\ell 4}=1$~TeV, so that the
fourth-generation leptons do not affect the decay branching ratios of the 
Higgs boson.  In both scenarios,
we choose the masses of the quarks to be those of the second scenario in Ref.~\cite{abf}, that is, we
set the mass of the fourth-generation down-type quark to be $m_{d4}=400$~GeV and 
the mass of the fourth-generation up-type quark to be 
$m_{u4}=m_{d4}+50~{\rm GeV}+10\log\left(m_H/115~{\rm GeV}\right){\rm GeV}$.  The other mass spectrum of
Ref.~\cite{abf} chooses $m_{d4}=300$~GeV, resulting in slightly larger predictions for 
$\sigma(gg\rightarrow H)$.
We use the next-to-next-to-leading order (NNLO) production cross section calculation of Ref.~\cite{abf}, which builds on the NNLO
SM calculations of Refs.~\cite{harlander1,melnikov1,ravindran1,anastasiou,bucherer1,spira1,nnloggh,aglietti},
the results of which are also listed in Ref.~\cite{EPAPS}.

The CDF and D0 Collaborations have searched for the SM Higgs boson in the
decay $H\rightarrow W^+W^-$ using all SM production processes:
$gg\rightarrow H$, $qq\rightarrow WH$, $qq\rightarrow ZH$, and vector-boson fusion (VBF)~\cite{cdfwwprl,d0wwprl,tevwwprl}.  
The results of these searches for the SM Higgs boson cannot be used directly to constrain
fourth-generation models, as the $ggH$ coupling is
enhanced but the $WWH$ and $ZZH$ couplings are not, and the
signal acceptances and the backgrounds in the multiple analysis channels
differ for the various production modes.  Therefore, these searches
rely on the SM to predict the ratios of the production rates of the 
$gg\rightarrow H$, $WH$, $ZH$, and VBF signals.  
Previous external analyses have used the Tevatron's SM Higgs boson search
results to constrain fourth-generation models, incorrectly
arguing that the $WH$, $ZH$, and VBF production rates are not significant, thus obtaining
only approximate results.  Furthermore, the SM results~\cite{cdfwwprl,d0wwprl,tevwwprl}
extend only up to $m_H$ of 200~GeV.
This paper addresses both of these issues by placing limits on $\sigma(gg\rightarrow H)\times 
\mathcal{B}(H\rightarrow W^+W^-)$ up to $m_H=300$~GeV.

Previously, the CDF and D0 collaborations have published searches for
the process $gg\rightarrow H\rightarrow W^+W^-$, also 
neglecting the $WH$, $ZH$, and VBF signal
contributions~\cite{D0GGH,CDFGGH}.  The D0 search includes a
fourth-generation interpretation.  Here we update these searches with
those using 4.8~fb$^{-1}$ from CDF~\cite{cdfwwprl} and 5.4~fb$^{-1}$
from D0~\cite{d0wwprl}. 
We present new limits on
$\sigma(gg\rightarrow H)\times \mathcal{B}(H\rightarrow W^+W^-)$ in which the $gg\rightarrow H$ 
production mechanism is considered as the unique signal source.  These limits
are compared to models for Higgs boson production in which the $ggH$
coupling is enhanced by the presence of a single additional generation
of fermions.  In this comparison, the decay branching ratios
of the Higgs boson are also modified to reflect changes
due to the fourth generation relative to the SM prediction.  
While the decays of the heavy quarks and leptons may include $W$ bosons
in the final state, we do not include these as additional sources of signal.
The branching ratios for $H\rightarrow W^+W^-$ are calculated using {\sc hdecay}~\cite{hdecay}
modified to include fourth-generation fermions~\cite{g4_hdecay}.
The modified Higgs branching ratio to $W^+W^-$ is multiplied by
the cross section~\cite{abf} to predict the fourth-generation enhanced
$gg\rightarrow H \rightarrow W^+W^-$ production rate.
 
The event selections are similar for the corresponding CDF and D0
analyses.  Both collaborations select events with large \met~and two
oppositely charged, isolated leptons, targeting the $H\rightarrow
W^+W^-$ signal in which both $W$ bosons decay leptonically.  The D0
analysis classifies events in three channels defined by the number of
charged leptons ($e$ or $\mu$), $e^+e^-$, $e^\pm \mu^\mp$, and
$\mu^+\mu^-$ and no classification based upon jet multiplicity.
The CDF analysis separates opposite-sign candidate
events into five non-overlapping channels.  Events are classified by
their jet multiplicity (0, 1, or $\ge$ 2), and the 0 and 1 jet channels
are further divided according to whether both leptons are in the
central part of the detector or if either lepton is in the forward part of the detector.  Two changes
have been made in the D0 event selection from the analysis presented
in Ref. \cite{d0wwprl}. For higher Higgs boson masses ($m_H > 200$~GeV),
the dilepton azimuthal-opening angle distribution is no longer peaked at low values ($\Delta\phi(\ell,\ell) <$ 1).
Therefore, to enhance the signal acceptance for large $m_H$, the requirement on the dilepton
azimuthal-opening angle [$\Delta\phi(\ell,\ell)$] has been removed for
$e^{\pm}\mu^{\mp}$ candidate events and relaxed to
$\Delta\phi(\ell,\ell) < 2.5$ in the $e^+e^-$ and $\mu^+\mu^-$
candidate events.
In addition, a requirement on the $\phi$-opening angle between the leading
muon and the missing transverse energy, $\Delta\phi(\mu,\met) > 0.5 $,
has been included to remove additional background in a signal-free
region.  The predicted contributions from the different background 
processes are compared with the numbers of events observed in data for the CDF and D0 analyses in Tables~\ref{tab:cdf_yields} and~\ref{tab:d0_yields}, respectively.

\begin{table*}[htb]
\caption{\label{tab:cdf_yields} Expected and observed event yields in
the 0-jet exclusive, 1-jet exclusive, and 2-jet inclusive samples at final selection for the CDF analysis summed across all lepton categories. The systematic uncertainty is shown for all samples.  The signal expectation is given for the low-mass fourth-generation scenario with a SM Higgs mass of 200~GeV with a predicted $\sigma(gg\rightarrow H)\times BR(H\rightarrow W^+W^-)$ of 1.02 pb.}
%\begin{ruledtabular}
\begin{tabular}{c |r r l|r r l|r r l}
\hline\hline
CDF Run II &  \\
$\int \mathcal{L} = 4.8 \; \rm{fb}^{-1}$&\multicolumn{3}{c}{0-jet}&\multicolumn{3}{c}{1-jet}&\multicolumn{3}{c}{$\ge 2$-jets}\\
\hline
\hline
$Z/\gamma^*\to \ell^+\ell^-$  & 128 & $\pm$ & 30& 133 & $\pm$ & 42& 51 & $\pm$ & 17\\
$t\bar t$ & 1.99 & $\pm$ & 0.31& 48.4 & $\pm$ & 7.6& 145 & $\pm$ & 24\\
$WW$ & 447 & $\pm$ & 48& 121 & $\pm$ & 13& 25.6 & $\pm$ & 5.8\\
$WZ$ & 19.7 & $\pm$ & 2.7& 20.0 & $\pm$ & 2.7& 5.30 & $\pm$ & 0.73\\
$ZZ$ & 29.9 & $\pm$ & 4.1& 8.0 & $\pm$ & 1.1& 2.36 & $\pm$ & 0.32\\
$W+{\rm jets}$ & 154 & $\pm$ & 37& 59 & $\pm$ & 15& 21.9 & $\pm$ & 5.9\\
$W\gamma$ & 112 & $\pm$ & 19& 16.2 & $\pm$ & 3.6& 2.72 & $\pm$ & 0.67\\
\hline
{ Total Background} & 893 & $\pm$ & 79& 406 & $\pm$ & 52& 254 & $\pm$ & 33\\
\hline
$gg\rightarrow H$ ($M_H = 200~\mathrm{GeV}$) & 35.2 & $\pm$ & 5.0& 20.2 & $\pm$ & 5.1& 8.5 & $\pm$ & 5.1\\
\hline\hline
{ Data} & \multicolumn{3}{c}{950}& \multicolumn{3}{c}{393}& \multicolumn{3}{c}{224}\\
\hline\hline
\end{tabular}
%\end{ruledtabular}
\end{table*}

\begin{table*}[hbt]
\begin{center}
\caption{\label{tab:d0_yields} Expected and observed event yields in
each channel at the final selection for the D0 analysis summed across all jet multiplicities. The systematic uncertainty after fitting is shown for all samples at final selection.  The signal expectation is given for the low-mass fourth-generation scenario with a SM Higgs mass of 200~GeV with a predicted $\sigma(gg\rightarrow H)\times BR(H\rightarrow W^+W^-)$ of 1.02 pb.}

\begin{tabular}{c|r@{ $\,\pm \,$ }l|r@{ $\,\pm \,$ }l|r@{ $\,\pm \,$ }l}
\hline \hline
D0 Run II \\
            $\int \mathcal{L} = 5.4 \; \rm{fb}^{-1}$   & \multicolumn{2}{c}{$e^{\pm}\mu^{\mp}$}&
                \multicolumn{2}{c}{$e^+e^-$}          &
                \multicolumn{2}{c}{$\mu^+\mu^-$} \\
\hline\hline
$Z/\gamma^*\to e^+e^-$       & \multicolumn{2}{c|}{$<0.1$} & 370 & 24 &     \multicolumn{2}{c}{$-$}  \\
$Z/\gamma^*\to \mu^+\mu^-$     & 7.0&0.1    &      \multicolumn{2}{c|}{$-$}            & 2056& 58\\
$Z/\gamma^*\to \tau^+\tau^-$  & 28.0&0.2       & 0.8 &0.1 &  6.9 & 0.6\\
$\ttbar$          & 176 & 15      & 58.9&5.5      & 74.9 & 6.8 \\
$WW$           & 304&18   & 102 &7.3     & 145 & 11 \\
$WZ$           & 13.4&0.2  & 18.1 &1.0    & 31.4&2.0 \\
$ZZ$           & 1.1&0.1 & 15.2 &0.9   & 26.9&1.7 \\
$W+{\rm jets}/\gamma$         & 156 &12    & 154&14   & 118 & 13.7 \\
Multijet       & 10.4&2.5    & 1.4&0.1      & 72.7&13.7 \\
\hline
{ Total Background}& 696 & 26 & 720 &32   & 2532 & 58 \\
\hline
$gg\rightarrow H$ ($M_H = 200~\mathrm{GeV}$) & 36.5 & 5.4   & 15.8 &2.2    & 19.0 & 2.9 \\
\hline\hline
{ Data }&  \multicolumn{2}{c}{~684}              & \multicolumn{2}{c}{~719}       & \multicolumn{2}{c}{2516} \\
\hline\hline

\end{tabular}
\end{center}
\end{table*}

The presence of neutrinos in the final state prevents event-by-event
reconstruction of the Higgs boson mass and thus other variables are used
for separating the signal from the background.  For example, the
angle $\Delta\phi(\ell,\ell)$ in signal events is smaller on average
than that in background events, the missing transverse momentum is
larger, and the total transverse energy of the jets is lower.  In
these analyses, the final discriminants are neural-network (NN)~\cite{cdfnn,d0nn} outputs
based on several kinematic variables.  For CDF, the list of network inputs includes likelihood ratio discriminant variables constructed from matrix-element probabilities~\cite{cdfwwprl}.

\begin{figure*}
 \begin{center}
 \includegraphics[width=0.3\textwidth]{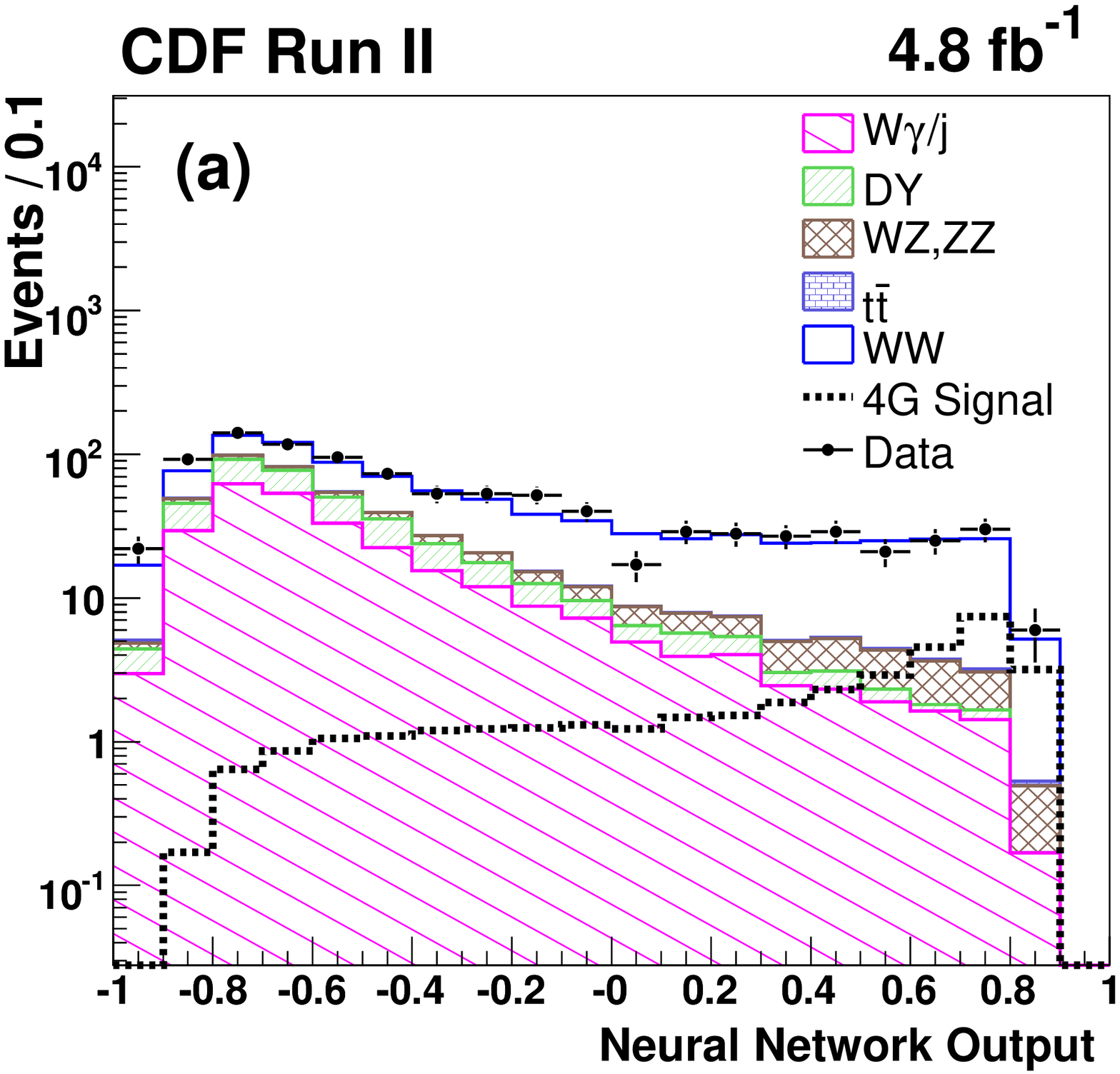} 
 \includegraphics[width=0.3\textwidth]{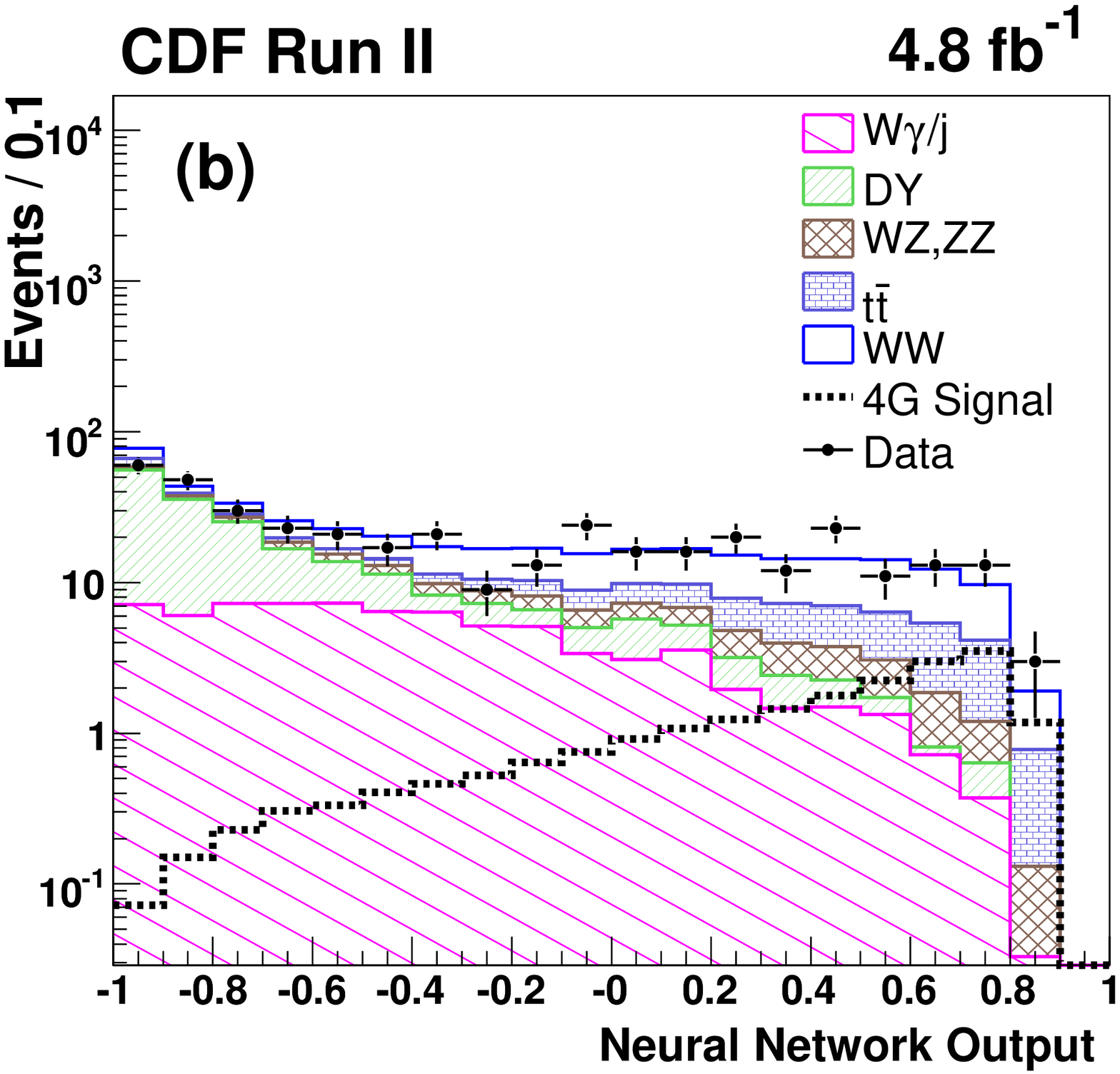} 
 \includegraphics[width=0.3\textwidth]{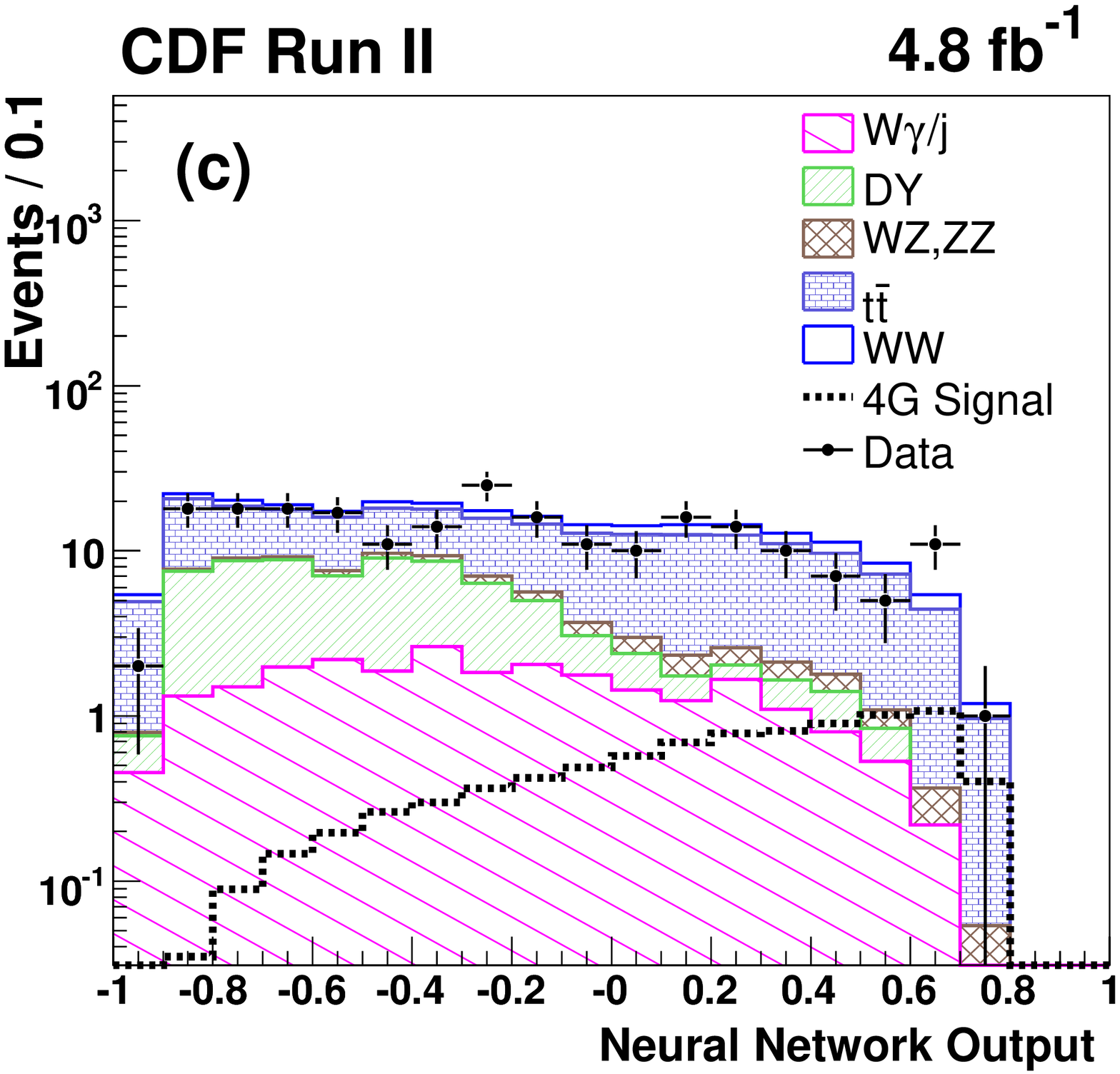} 
 \end{center}
 \caption{
 \label{fig:cdf_nndist}
Distributions of the neural network outputs for the search for a 
Higgs boson of mass $m_H=200$~GeV, from CDF.  The data are shown as points with uncertainty bars, and the background
predictions are shown stacked.  The figures show the distributions for events with (a) zero, (b) one, and
(c) two or more identified jets, respectively.  The distributions are summed over lepton categories.  The fourth-generation signal, 
normalized to the prediction of the low-mass scenario, is shown not stacked.}
 \end{figure*}

\begin{figure*}
 \begin{center}
 \includegraphics[width=0.3\textwidth]{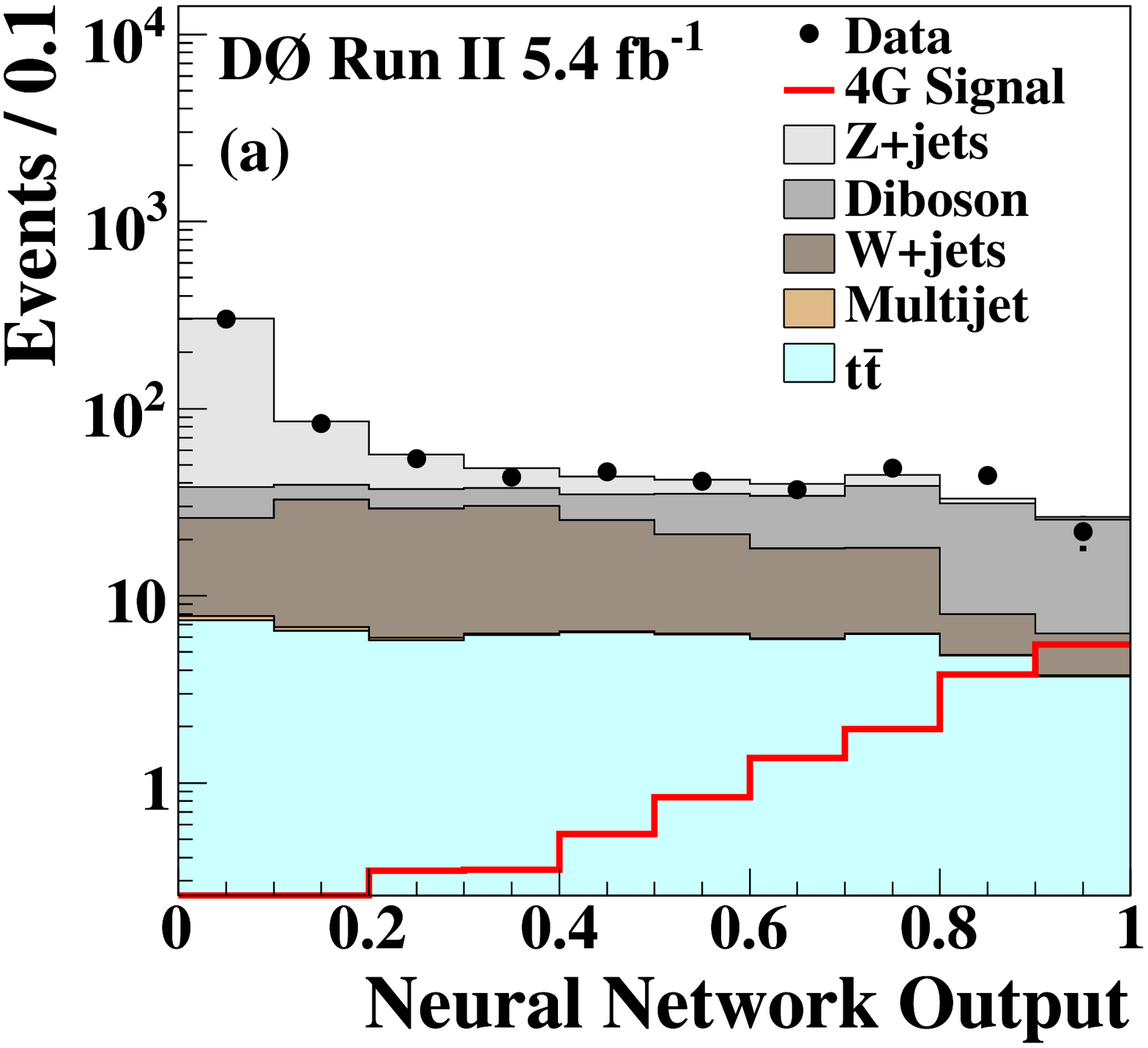}
 \includegraphics[width=0.3\textwidth]{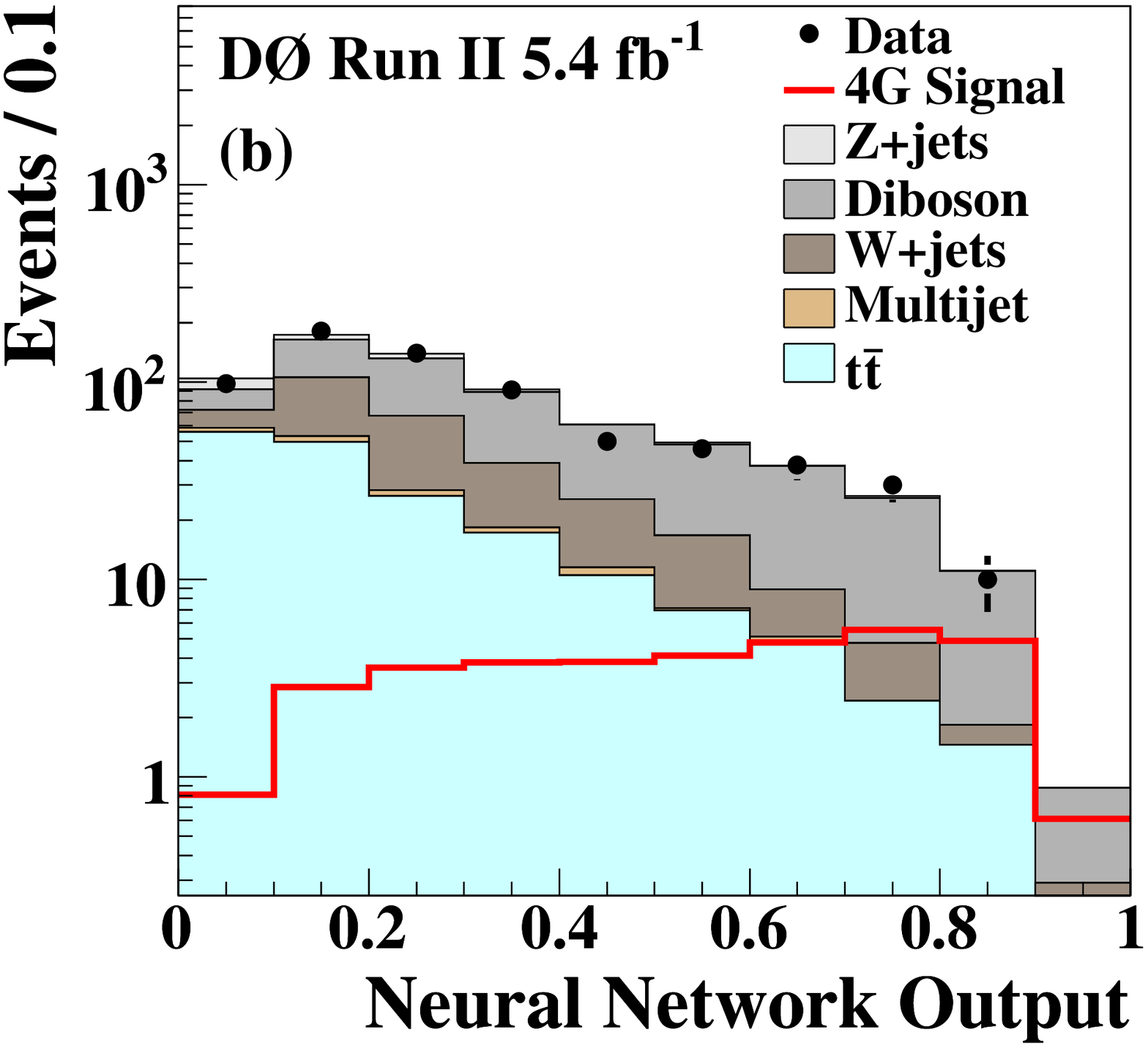}
 \includegraphics[width=0.3\textwidth]{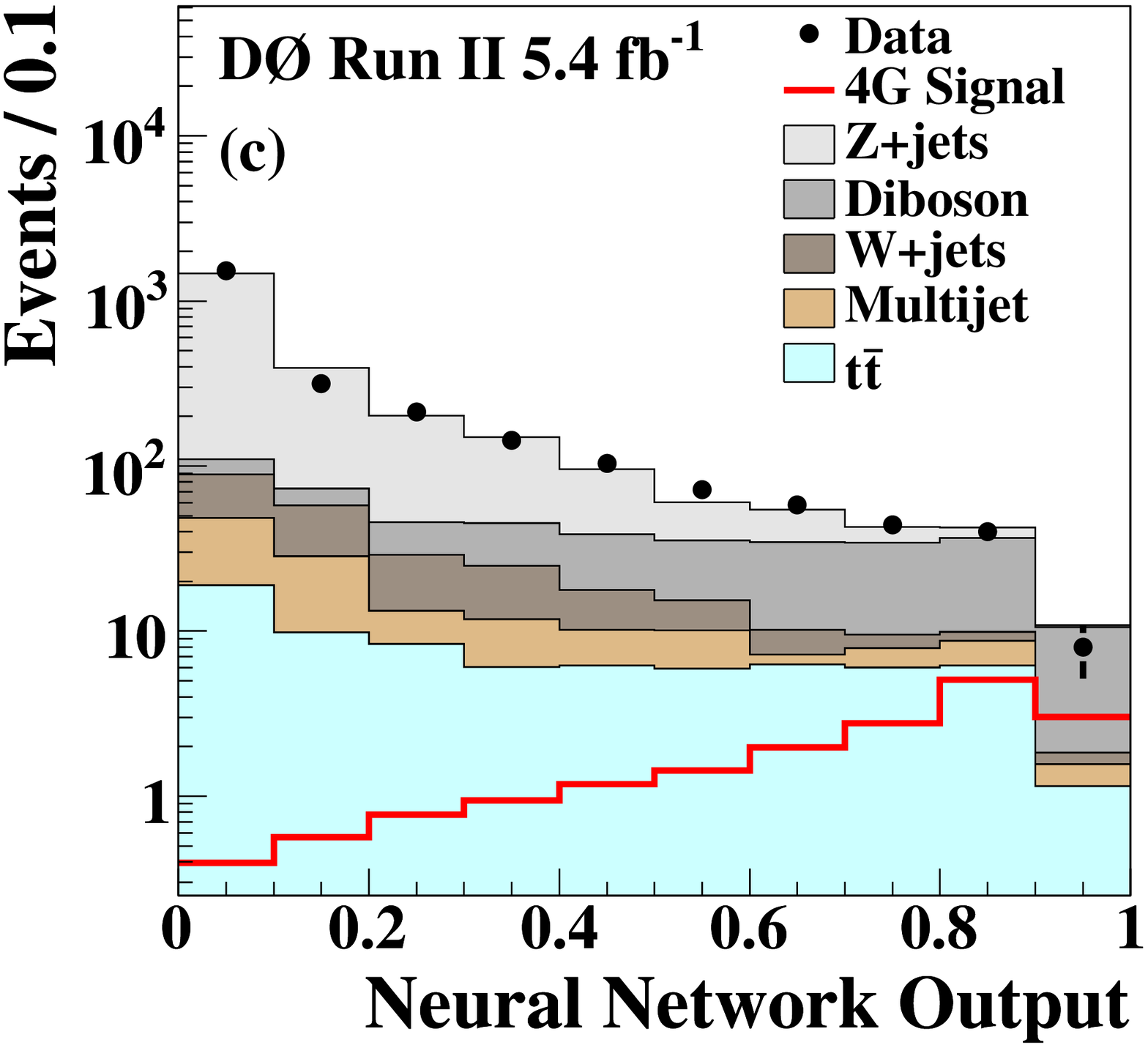}
 \end{center}
 \caption{
 \label{fig:d0_nndist}
Distributions of the neural network outputs for the 
search for a Higgs boson of mass $m_H=200$~GeV, from D0 summed over all jet multiplicities.  (a) shows the distribution
for the di-electron selection, (b) shows the distribution for the electron-muon selection, and
(c) shows the distribution for the di-muon selection.
The data are shown as points with uncertainty bars, and the background
predictions are shown stacked.  The background uncertainty
is the post-fit systematic uncertainty.  The fourth-generation signal, normalized to the prediction
of the low-mass scenario, is shown not stacked. }
 \end{figure*}

Both CDF and D0 have extended their searches to the range $110<m_H<300$~GeV.
Separate neural networks are trained to distinguish the $gg\rightarrow H$ 
signal from the backgrounds for each of the test masses, which are separated by increments of 5 or 10 GeV.  Distributions of
the network outputs for CDF and D0 are shown in
Figs.~\ref{fig:cdf_nndist} and \ref{fig:d0_nndist}, comparing the data
with predictions for a Higgs boson of mass
$m_H=200$~GeV.  Because the background composition and the signal kinematics are functions of
the number of jets in the event, the CDF NN output distributions are shown separately for
0, 1, and 2 or more jets, summed over lepton categories.
For D0, as the detector response is different for electrons and muons,
the NN distributions are shown separately
for $e^+e^-$, $e^\pm \mu^\mp$, and $\mu^+\mu^-$ selections.

The details of the signal and background estimations and the
systematic uncertainties are provided in
Refs.~\cite{cdfwwprl,d0wwprl,tevwwprl}.  We set
limits on $\sigma(gg\rightarrow H)\times \mathcal{B}(H\rightarrow W^+W^-)$ as a
function of $m_H$. 
We use the same two statistical methods employed in
Ref.~\cite{tevwwprl}, namely the modified frequentist (CL$_{\rm s}$) and Bayesian techniques in
order to study the consistency of the results.  Each method is
applied at each test mass to calculate an observed upper limit on
$\sigma(gg\rightarrow H)\times \mathcal{B}(H\rightarrow W^+W^-)$.
Pseudo-experiments drawn from systematically varied background-only
predictions are used to compute the limits we expect to obtain in the
absence of a signal.  We present both the Bayesian and CL$_{\rm s}$
observed and expected limits in Ref.~\cite{EPAPS}.   The limits calculated with
the two methods agree within 6\% for all Higgs boson mass hypotheses.
Correlated systematic
uncertainties are treated in the same way as they are in
Ref.~\cite{tevwwprl}.  The sources of correlated uncertainty between
CDF and D0 are the total inelastic $p\bar{p}$ cross section used in the luminosity measurement, the SM
diboson background production cross sections ($WW$, $WZ$, and $ZZ$),
and the $t{\bar{t}}$ and single top quark production cross sections.
Instrumental effects such as trigger efficiencies, lepton identification
efficiencies and misidentification rates, and the jet energy scales
used by CDF and D0 remain uncorrelated.  To minimize the degrading effects of systematics on the search sensitivity, 
the signal and background contributions are fit to the data observations by maximizing a likelihood function over the systematic uncertainties for both the background-only and signal+background hypotheses~\cite{fitting}.  When setting limits on
$\sigma(gg\rightarrow H)\times \mathcal{B}(H\rightarrow W^+W^-)$, we do not include
the theoretical uncertainty on the prediction of $\sigma(gg\rightarrow H)\times \mathcal{B}(H\rightarrow W^+W^-)$
in the fourth-generation models since these limits are independent of the predictions.
When setting limits on $m_H$ in the context of fourth-generation models, however,
we include the uncertainties on the theoretical predictions as described below.

\begin{figure}
 \begin{center}
 \includegraphics[width=0.9\columnwidth]{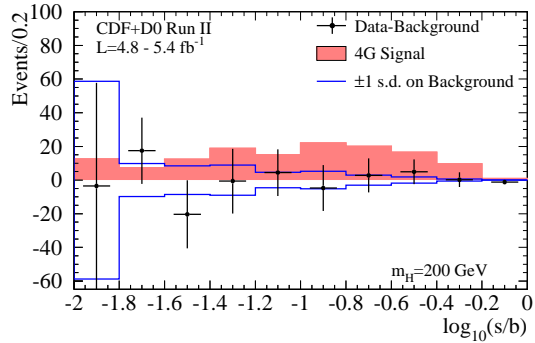}
 \end{center}
 \caption{
 \label{fig:tev_bg_subtracted} Background-subtracted data
distribution for the discriminant histograms, summed for bins of $s/b$,
for the $m_H$ = 200~GeV combined search. The background is fitted to the data under the background-only hypothesis, and the uncertainty on the background is the post-fit systematic uncertainty. The signal,
which is normalized to the low-mass fourth-generation SM
expectation, is shown with a filled histogram. The uncertainties shown
on the background-subtracted data points are the square roots of the
post-fit background predictions in each bin, representing the expected
statistical uncertainty on the data.
}
 \end{figure}

Before computing the cross-section limits, we investigate the properties of the signal and background
predictions in each bin of the analyses, as well as those of the observed data.
Because there are many
channels to combine, we represent the data in a compact form by sorting the bins of each analysis
by their signal-to-background ratio $s/b$, where $s$ and $b$ are the number of signal and background events, repetitively.  The predictions and observations in bins of similar $s/b$ are then collected.
For the $m_H=200$~GeV search, the background subtracted data distribution 
compared with the signal prediction can be seen in Fig.~\ref{fig:tev_bg_subtracted}.  The background
used and its uncertainties are shown after fitting to the data.
 No significant excess is observed in the data, and the theory predicts a measurable excess over
the background.

\begin{figure*}
 \begin{center}
 \includegraphics[width=0.45\textwidth]{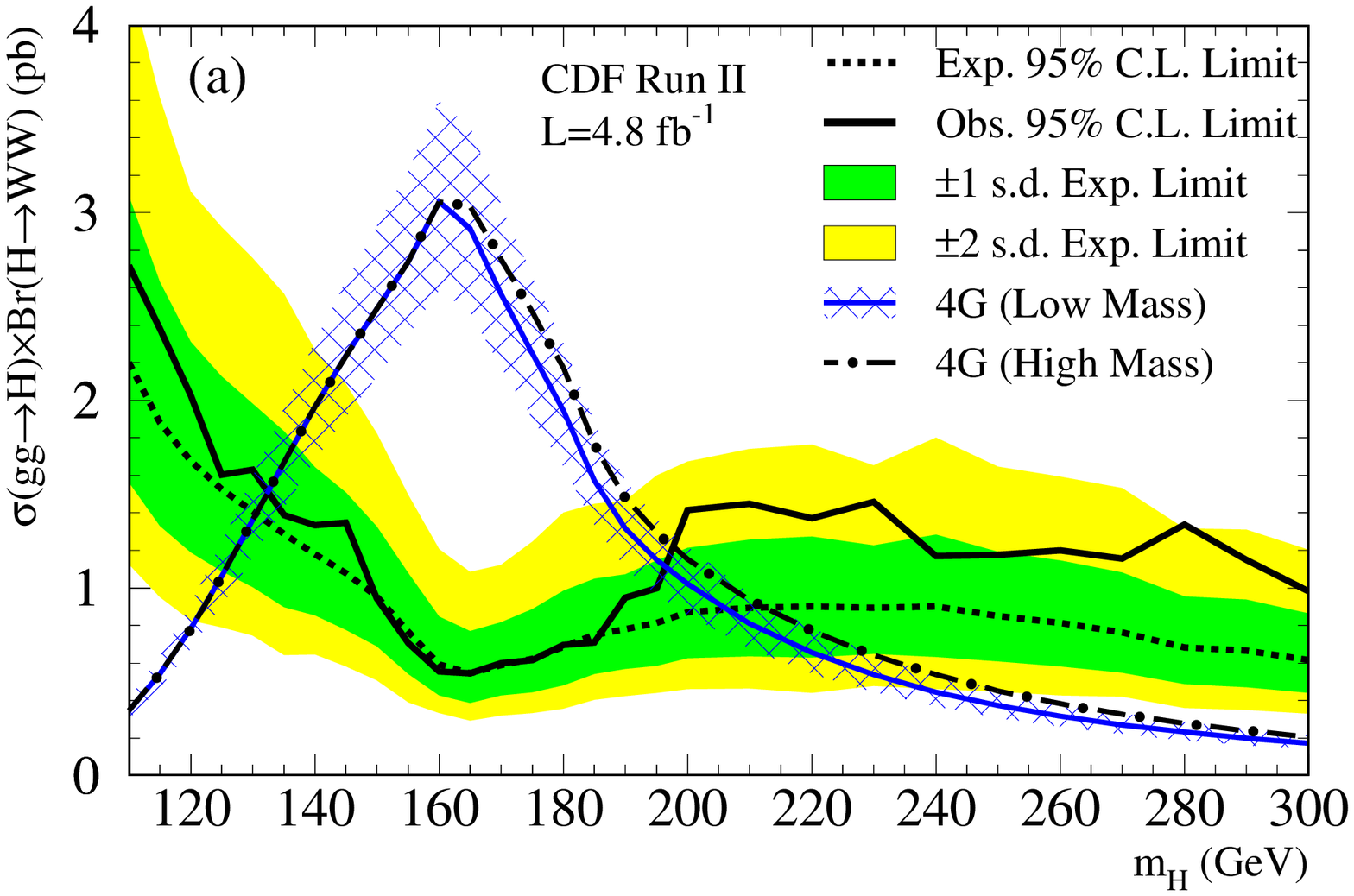}\includegraphics[width=0.45\textwidth]{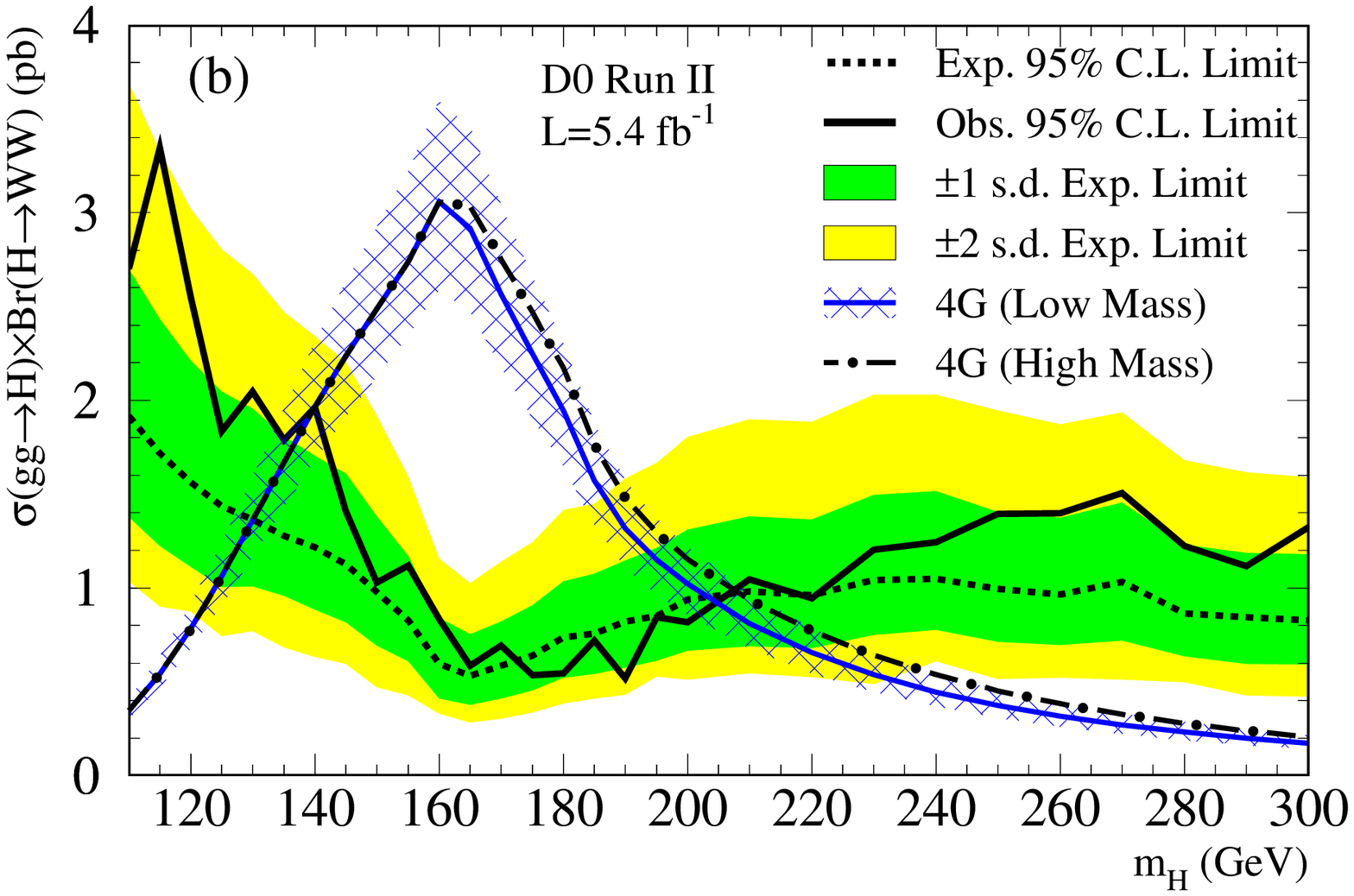}
 \includegraphics[width=0.45\textwidth]{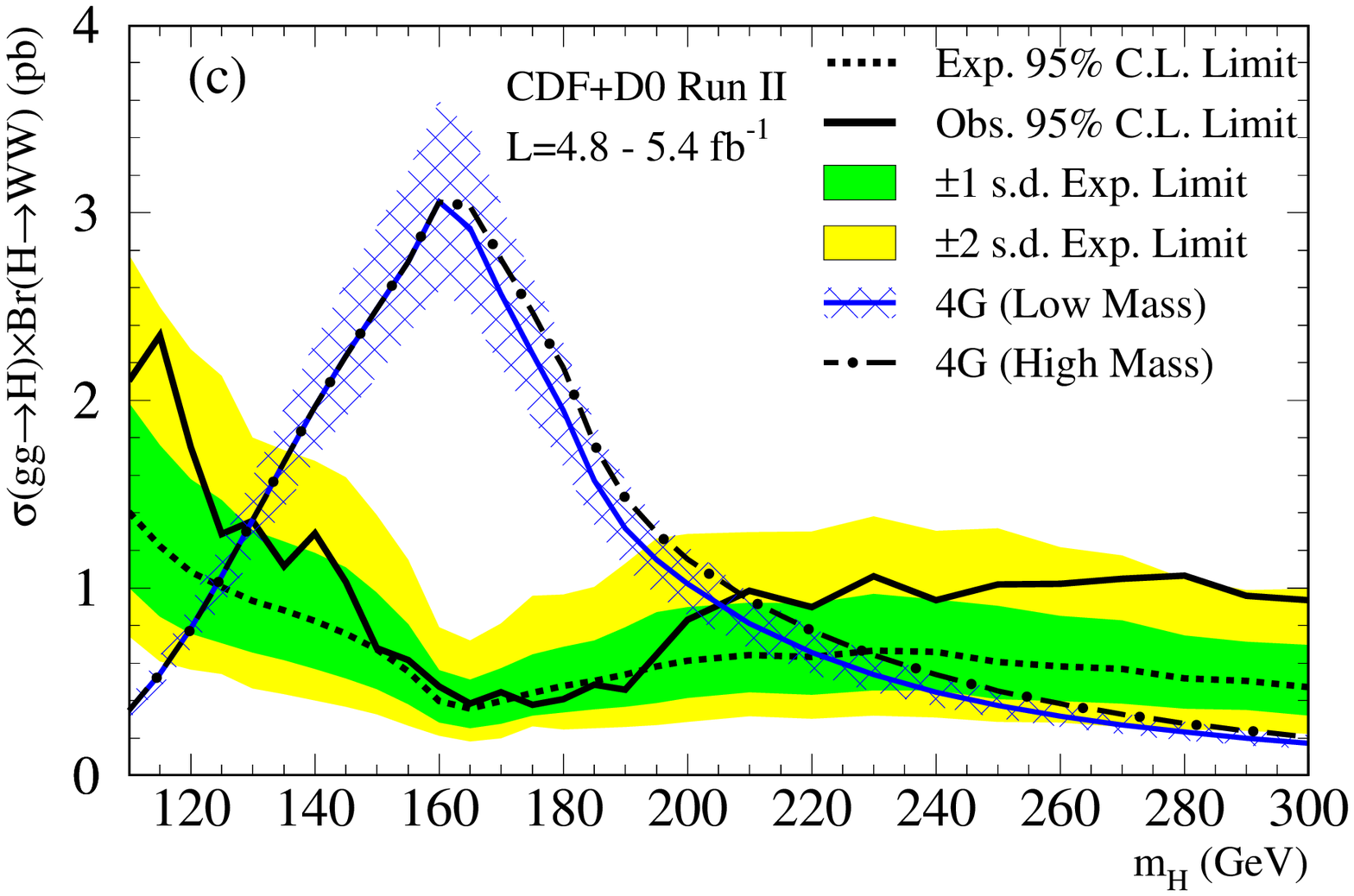}\includegraphics[width=0.45\textwidth]{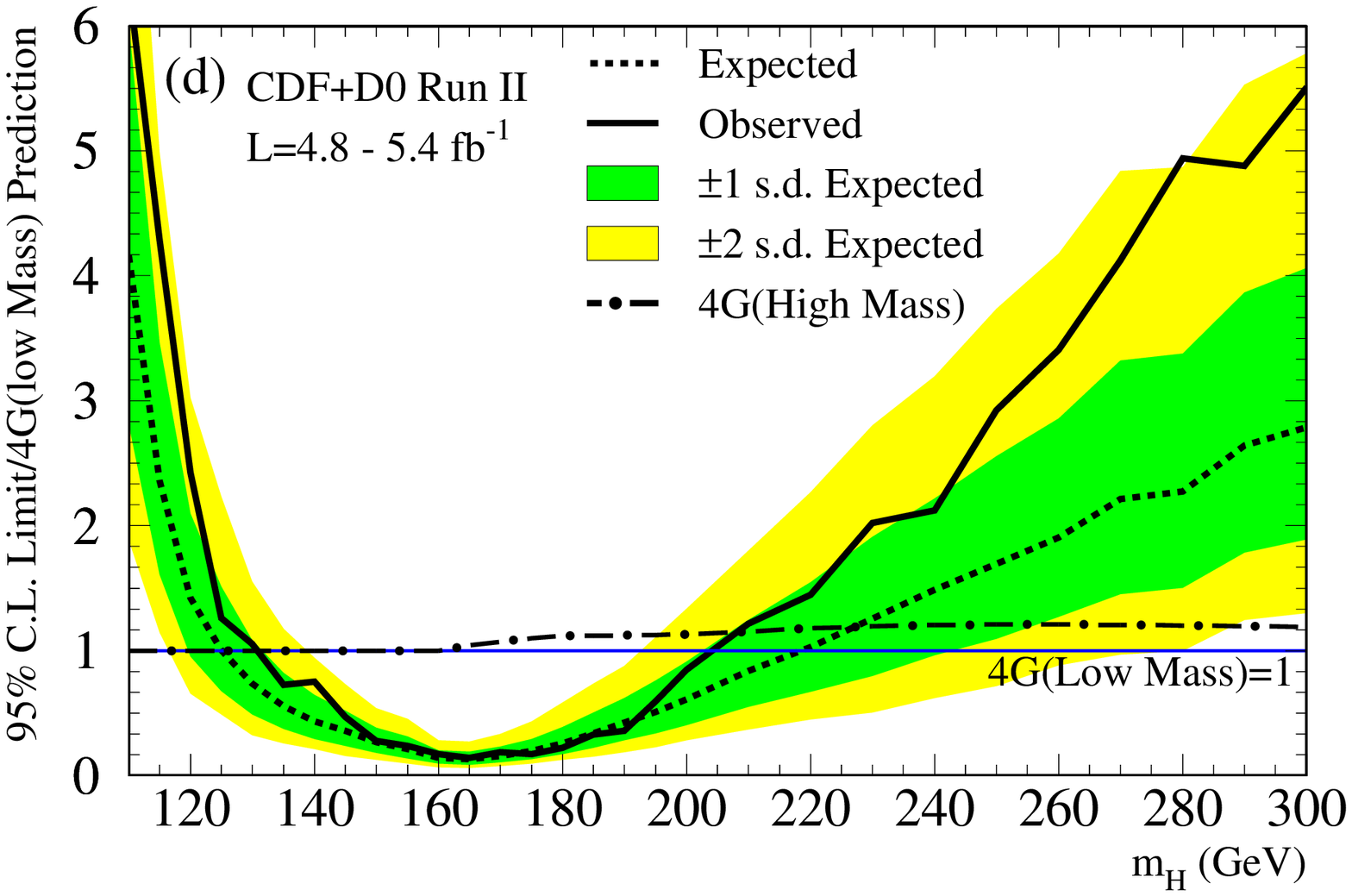}
 \end{center}
 \caption{
 \label{fig:limits}  The CDF, D0, and combined  observed (solid black lines) and median expected (dashed
black lines) 95\% C.L. upper limits on $\sigma(gg\rightarrow H)\times \mathcal{B}(H\rightarrow W^+W^-)$ are shown
in figures (a) through (c).  The shaded bands indicate the $\pm 1$~standard deviation (s.d.) and $\pm 2$~s.d. intervals 
on the distribution of the 
limits that are expected if a Higgs boson signal is not present.  Also shown on each graph is the prediction
for a fourth-generation model in the low-mass and high-mass scenarios, 4G (Low mass) and 4G (High mass) respectively.
The hatched areas indicate the theoretical uncertainty from PDF and scale uncertainties.  
The lighter curves show the high-mass theoretical prediction.  Figure (d) shows the 95\% C.L.  combined
limit relative to the low-mass theoretical prediction, where the uncertainties in the signal prediction
are included in the limit.  Also shown in Figure (d) is the prediction of the signal rate
in the high-mass scenario, divided by that of the low-mass scenario.
}
 \end{figure*}

The separate limits on $\sigma(gg\rightarrow H)\times \mathcal{B}(H\rightarrow W^+W^-)$ from
CDF and D0 are shown in Figs.~\ref{fig:limits}(a)
and~\ref{fig:limits}(b), respectively.  Since CDF separates the
different jet categories into separate channels, theoretical uncertainties on the relative contributions
of the $gg\rightarrow H$ signal in the separate jet channels~\cite{ag} 
are included in the same way as signal acceptance uncertainties.
The combined limits on $\sigma(gg\rightarrow H)\times \mathcal{B}(H\rightarrow W^+W^-)$
are shown in Fig.~\ref{fig:limits}(c) along with the fourth-generation
theory predictions for the high-mass and low-mass scenarios.  The
uncertainty bands shown on the low-mass theoretical prediction are the
sum in quadrature of the MSTW 2008 \cite{mstw1} 90\% C.L. parton
distribution function (PDF) uncertainties and the factorization and
renormalization scale uncertainties from Table~1 of Ref.~\cite{abf},
which are also reported Ref.~\cite{EPAPS}, giving a total uncertainty
of 15\% for $m_H = $ 160~GeV.  The scale uncertainties are determined
by recalculating the cross sections with the scale multiplied by
factors of $1/2$ and $2$.  The scale uncertainties are independent of
$m_H$ and are similar to the uncertainties for SM
$\sigma(gg\rightarrow H)$ predictions~\cite{anastasiou,grazzini}.  The
PDF uncertainties, however, grow with increasing $m_H$, as higher-$x$
gluons are required to produce more massive Higgs bosons.

In order to set limits on $m_H$ in these two scenarios, we perform a second
combination, including the uncertainties on the theoretical predictions of
$\sigma(gg\rightarrow H)\times \mathcal{B}(H\rightarrow W^+W^-)$ due to scale
and PDF uncertainties at each value of
$m_H$ tested.  The resulting limits are computed relative to the model
prediction, and are shown in Fig.~\ref{fig:limits}(d) for the low-mass
scenario, which gives the smaller excluded range of $m_H$.  In this scenario,
we exclude at the 95\% C.L. a SM-like Higgs boson with a mass
in the range 131 -- 204~GeV.  Using the median  
limits on $\sigma(gg\rightarrow H)\times \mathcal{B}(H\rightarrow W^+W^-)$, expected
in the absence of a signal, to quantify the sensitivity, we expect to exclude the mass range 125 -- 218~GeV.
In the high-mass scenario, which predicts a larger $\mathcal{B}(H\rightarrow W^+W^-)$ 
at high $m_H$ than that predicted in the low-mass scenario,
we exclude at the 95\% C.L. the mass range 131 -- 208~GeV and expect to exclude the mass
range 125 -- 227~GeV.

In summary, we presented a combination of CDF and D0 searches for the $gg\rightarrow H\rightarrow W^+W^-$
process and set an upper limit on $\sigma(gg\rightarrow H)\times \mathcal{B}(H\rightarrow W^+W^-)$ as a function
of $m_H$.  We compared these limits with the prediction of the minimal SM with a sequential
fourth generation of heavy fermions added on, and excluded at the 95\% C.L. the Higgs boson mass range $131<m_H<204$~GeV, with an expected
excluded range of 125 -- 218 GeV.

\begin{center}
{\bf Acknowledgements}
\end{center}

We thank the Fermilab staff and the technical staffs of the
participating institutions for their vital contributions. 
This work was supported by  
DOE and NSF (USA),
CONICET and UBACyT (Argentina), 
CNPq, FAPERJ, FAPESP and FUNDUNESP (Brazil),
CRC Program, CFI, NSERC and WestGrid Project (Canada),
CAS and CNSF (China),
Colciencias (Colombia),
MSMT and GACR (Czech Republic),
Academy of Finland (Finland),
CEA and CNRS/IN2P3 (France),
BMBF and DFG (Germany),
Ministry of Education, Culture, Sports, Science and Technology (Japan), 
World Class University Program, National Research Foundation (Korea),
KRF and KOSEF (Korea),
DAE and DST (India),
SFI (Ireland),
INFN (Italy),
CONACyT (Mexico),
NSC(Republic of China),
FASI, Rosatom and RFBR (Russia),
Slovak R\&D Agency (Slovakia), 
Ministerio de Ciencia e Innovaci\'{o}n, and Programa Consolider-Ingenio 2010 (Spain),
The Swedish Research Council (Sweden),
Swiss National Science Foundation (Switzerland), 
FOM (The Netherlands),
STFC and the Royal Society (UK),
and the A.P. Sloan Foundation (USA).

With visitors to CDF from 
$^{a\dag}$University of Massachusetts Amherst, Amherst, Massachusetts 01003;
$^{b\dag}$Universiteit Antwerpen, B-2610 Antwerp, Belgium;
$^{c\dag}$University of Bristol, Bristol BS8 1TL, United Kingdom;
$^{d\dag}$Chinese Academy of Sciences, Beijing 100864, China;
$^{e\dag}$Istituto Nazionale di Fisica Nucleare, Sezione di Cagliari, 09042 Monserrato (Cagliari), Italy;
$^{f\dag}$University of California Irvine, Irvine, CA  92697;
$^{g\dag}$University of California Santa Cruz, Santa Cruz, CA  95064;
$^{h\dag}$Cornell University, Ithaca, NY  14853;
$^{i\dag}$University of Cyprus, Nicosia CY-1678, Cyprus;
$^{j\dag}$University College Dublin, Dublin 4, Ireland;
$^{k\dag}$University of Edinburgh, Edinburgh EH9 3JZ, United Kingdom;
$^{l\dag}$University of Fukui, Fukui City, Fukui Prefecture, Japan 910-0017;
$^{m\dag}$Kinki University, Higashi-Osaka City, Japan 577-8502;
$^{n\dag}$Universidad Iberoamericana, Mexico D.F., Mexico;
$^{o\dag}$University of Iowa, Iowa City, IA  52242;
$^{p\dag}$Iowa State University, Ames, IA 50011;  
$^{q\dag}$Kansas State University, Manhattan, KS 66506;
$^{r\dag}$Queen Mary, University of London, London, E1 4NS, England;
$^{s\dag}$University of Manchester, Manchester M13 9PL, England;
$^{t\dag}$Muons, Inc., Batavia, IL 60510;
$^{u\dag}$Nagasaki Institute of Applied Science, Nagasaki, Japan;
$^{v\dag}$University of Notre Dame, Notre Dame, IN 46556;
$^{w\dag}$Obninsk State University, Obninsk, Russia;
$^{x\dag}$University de Oviedo, E-33007 Oviedo, Spain;
$^{y\dag}$Texas Tech University, Lubbock, TX  79609;
$^{z\dag}$IFIC(CSIC-Universitat de Valencia), 56071 Valencia, Spain;
$^{aa\dag}$Universidad Tecnica Federico Santa Maria, 110v Valparaiso, Chile;
$^{bb\dag}$University of Virginia, Charlottesville, VA  22906;
$^{cc\dag}$Bergische Universit\"at Wuppertal, 42097 Wuppertal, Germany;
$^{dd\dag}$Yarmouk University, Irbid 211-63, Jordan; and
$^{ee\dag}$On leave from J.~Stefan Institute, Ljubljana, Slovenia,
and with visitors to D0 from
$^{ff\ddag}$Augustana College, Sioux Falls, SD, 61201;
$^{gg\ddag}$The University of Liverpool, Liverpool, UK;
$^{hh\ddag}$SLAC, Menlo Park, CA, 94025;
$^{ii\ddag}$ICREA/IFAE, Barcelona, Spain;
$^{jj\ddag}$Centro de Investigacion en Computacion - IPN, Mexico City, Mexico;
$^{kk\ddag}$ECFM, Universidad Autonoma de Sinaloa, Culiac\'an, Mexico; and
$^{ll\ddag}$Universit{\"a}t Bern, Bern, Switzerland.
\noaffiliation

\begin{table*}
\begin{center}
{\Large{\bf Auxiliary material}}
\end{center}
\end{table*}

\setcounter{figure}{0}
\setcounter{table}{0}

\begin{table*}[htb]
\caption{\label{tab:cross_sections} The fourth-generation enhanced $\sigma(gg\rightarrow H)$ listed in fb for the low-mass scenario described in the text with $m_{d4} = 300$~GeV and $m_{d4} = 400$~GeV along with the $\mathcal{B}(H\rightarrow W^+W^-)$ for Higgs masses between 110 - 300~GeV.  The uncertainty on the predicted cross section from variations in the PDF and factorization and renormalization scale are also listed in percentage where these have been determined from the MSTW 2008 90\% C.L. uncertainty and by modification of  the scale by factors of $1/2$ and 2, respectively.}
\begin{ruledtabular}
\begin{tabular}{lcccccccc}
$m_H$&$\sigma(gg\rightarrow H)$ &$\sigma(gg\rightarrow H)$ &uncert.&uncert.&uncert.&uncert.& $BR(H\rightarrow W^+W^-)$&$BR(H\rightarrow W^+W^-)$\\
 (GeV)&$m_{d4}=300$~GeV&$m_{d4}=400$~GeV&PDF up(\%) &PDF down(\%) &$\mu$ up(\%) &$\mu$ down(\%) &$m_{d4}=300$~GeV&$m_{d4}=400$~GeV\\
\hline\hline
110& 12384  &12308 & 12 &-11 &12&-8& 0.028&0.028 \\
115 &10798  &10725  &12 &-11 &12&-8& 0.050&0.051\\
120 &9449.9 &9384.3 &12 &-11 &12&-8&0.083& 0.083\\
125 &8298.8 &8240.0 &12 &-12 &12&-8&0.13&0.13     \\
130 &7314.0 &7258.7 &12 &-12 &12&-8&0.19     &0.19  \\   
135 &6465.1 &6414.2 &12 &-12 &12&-8&0.26     &0.26    \\ 
140 &5731.4 &5684.1 &13 &-12 &12&-8&0.35     &0.35     \\
145 &5094.6 &5050.4 &13 &-12 &12&-8&0.44     &0.44     \\
150 &4540.5 &4498.5 &13 &-12 &12&-8&0.55   &  0.55     \\
155 &4055.6 &4017.6 &13 &-12 &12&-8&0.68&0.68 \\
160 &3630.2 &3595.1 &13 &-13 &12&-8&0.85 &0.85 \\
165 &3253.7 &3220.7 &14 &-13 &12&-8&0.91 &0.91 \\
170 &2924.1 &2893.2 &14 &-13 &12&-8&0.89 &0.88 \\
175 &2633.9 &2604.4 &14 &-13 &12&-8&0.86 &0.86 \\
180 &2376.7 &2348.9 &14 &-13 &12&-8&0.83 &0.83 \\
185 &2147.2 &2121.5 &15 &-13 &12&-8&0.74 &0.74 \\
190 &1943.9 &1919.7 &15 &-14 &12&-8&0.69 &0.69 \\
195 &1763.2 &1740.2 &15 &-14 &12&-8&0.66 &0.66 \\
200 &1601.8 &1580.0 &15 &-14 &12&-8&0.65 &0.65 \\
210 &1328.1 &1308.4 &16 &-14 &12&-8&0.62 & 0.62 \\
220 &1107.7 &1089.6 &16& -15 &12&-8&0.60 &0.60 \\
230 &928.61 &912.21 &17& -15 &12&-8&0.59 &0.59 \\
240 &782.52 &767.44 &17 &-15 &12&-8&0.58 &0.58 \\
250 &662.60 &648.81 &18& -16 &12&-8&0.58 &0.58 \\
260 &563.53 &550.90 &19 &-16 &12&-8&0.57 &0.58 \\
270 &481.49 &469.93 &19 &-16 &12&-8&0.57 &0.57 \\
280 &413.24 &402.68 &20 &-17 &12&-8&0.58 &0.58 \\
290 &356.39 &346.53 &21& -17 &12&-8&0.58 &0.58 \\
300 &308.70 &299.71 &21 &-17 &12&-8&0.58 &0.58 \\
\end{tabular}
\end{ruledtabular}
\end{table*}

\begin{table*}[htb]
\caption{\label{tab:limits} The observed and median expected 95\% C.L. upper limits on 
$\sigma(gg\rightarrow H)\times \mathcal{B}(H\rightarrow W^+W^-)$ for $m_H$ between
110~GeV and 300~GeV, obtained with the Bayesian and CL$_{\rm s}$ methods.  The ratio of the observed and expected Bayesian 95\% C.L. upper limits to the $\sigma(gg\rightarrow H)\times \mathcal{B}(H\rightarrow W^+W^-)$~predictions of the low-mass fourth generation scenario are listed.
Also listed are the $\sigma(gg\rightarrow H)\times \mathcal{B}(H\rightarrow W^+W^-)$~predictions of the low-mass and the high-mass fourth-generation scenarios discussed
in the text for a fourth-generation down-type quark mass of 400~GeV.  All limits and predictions are presented in pb.}
\begin{ruledtabular}
\begin{tabular}{lcccccccc}
 $m_H$       &  \multicolumn{4}{c}{Bayes}& \multicolumn{2}{c}{CL$_{\rm s}$}&  4$^{\rm{th}}$ Gen &  4$^{\rm{th}}$ Gen \\ 
 $\left[{\rm GeV}\right]$ &Obs. &Ratio Low Mass & Exp.& Ratio Low Mass   &   Obs.             & Exp.          & Low Mass    & High Mass\\
 &&(Obs./4Gen)&&(Exp./4Gen)&&&\\\hline
 110   &  2.10   &6.2 &    1.41       &4.2     &    2.07               &  1.40         &        0.34 &  0.35	 \\ 
 115   &  2.35    &4.4&    1.22 &2.3           &    2.34               &  1.25         &        0.54 &  0.54	 \\ 
 120   &  1.75    &2.2&    1.08         &1.4   &    1.77               &  1.15         &        0.78 &  0.78	 \\ 
 125   &  1.29    &1.2&    1.00           &0.94 &    1.25               &  1.05         &        1.06 &  1.06	 \\ 
 130   &  1.36    &1.0&    0.93 &0.68           &    1.41               &  0.98         &        1.36 &  1.36	 \\ 
 135   &  1.12    &0.67&    0.88      &0.53      &    1.14               &  0.90         &        1.67 &  1.67	 \\ 
 140   &  1.29    &0.66&    0.82&0.42            &    1.25               &  0.83         &        1.96 &  1.96	 \\ 
 145   &  1.03    &0.46&    0.76    &0.34        &    0.99               &  0.80         &        2.25 &  2.24	 \\ 
 150   &  0.68    &0.27&    0.67        &0.27    &    0.65               &  0.68         &        2.49 &  2.49	 \\ 
 155   &  0.62    &0.23&    0.56 &0.20           &    0.59               &  0.55         &        2.74 &  2.74	 \\ 
 160   &  0.47    &0.15&    0.40            &0.13&    0.47               &  0.40         &        3.06 &  3.06	 \\ 
 165   &  0.38    &0.13&    0.36 &0.12           &    0.37               &  0.36         &        2.92 &  3.03	 \\ 
 170   &  0.45    &0.18&    0.40            &0.16&    0.43               &  0.40         &        2.57 &  2.76	 \\ 
 175   &  0.38    &0.17&    0.44  &0.20          &    0.36               &  0.44         &        2.25 &  2.47    \\ 
 180   &  0.41    &0.21&    0.48            &0.24&    0.40               &  0.48         &        1.96 &  2.17	 \\ 
 185   &  0.48    &0.31&    0.50  &0.32          &    0.48               &  0.53         &        1.57 &  1.76	 \\ 
 190   &  0.46    &0.35&    0.54            &0.41&    0.46               &  0.55         &        1.32 &  1.48	 \\ 
 195   &  0.65    &0.57&    0.58&0.50            &    0.64               &  0.60         &        1.15 &  1.30	 \\ 
 200   &  0.83    &0.81&    0.61           &0.60 &    0.84               &  0.64         &        1.02 &  1.15	 \\ 
 210   &  0.98    &1.2&    0.64  &0.79          &    0.99               &  0.68         &        0.81 &  0.94    \\ 
 220   &  0.90    &1.4&    0.63            &0.97&    0.93               &  0.66         &        0.65 &  0.77	 \\ 
 230   &  1.06    &2.0&    0.67 &1.2           &    1.09               &  0.69         &        0.54 &  0.64	 \\ 
 240   &  0.93    &2.1&    0.66            &1.5&    0.90               &  0.67         &        0.45 &  0.54	 \\ 
 250   &  1.02    &2.8&    0.60 &1.6           &    1.02               &  0.63         &        0.37 &  0.45	 \\ 
 260   &  1.02    &3.2&    0.58            &1.8&    1.07               &  0.62         &        0.32 &  0.38    \\ 
 270   &  1.05    &3.9&    0.57  &2.1          &    1.07               &  0.60         &        0.27 &  0.33    \\ 
 280   &  1.07    &4.7&    0.52            &2.3&    1.08               &  0.53         &        0.23 &  0.28    \\ 
 290   &  0.96    &4.8&    0.50  &2.5          &    0.92               &  0.50         &        0.20 &  0.24    \\ 
 300   &  0.93    &5.5&    0.47        &2.8    &    0.91               &  0.48         &        0.17 &  0.21    \\ 
\end{tabular}
\end{ruledtabular}
\end{table*}
\end{document}